\documentclass[aps, reprint, superscriptaddress]{revtex4-2}
\usepackage[utf8]{inputenc}
\usepackage{graphicx}
\usepackage{amsmath, amssymb}
\usepackage{xcolor}
\usepackage{upgreek}
\usepackage{bm}
\usepackage{hyperref}

\newcommand{\D}{\mathrm{d}}
\newcommand{\ssb}{\mathbf{s}}
\newcommand{\Eb}{\mathbf{E}}
\newcommand{\Bb}{\mathbf{B}}
\newcommand{\vb}{\mathbf{v}}
\newcommand{\Omegab}{\bm{\Omega}}

\begin{document}

	\title{Pinching injection in wakefields for spin-polarized electron beams}

	\author{Lars Reichwein}
	\email{l.reichwein@fz-juelich.de}
     \affiliation{Peter Gr\"{u}nberg Institut (PGI-6), Forschungszentrum J\"{u}lich, 52425 J\"{u}lich, Germany}
     \affiliation{State Key Laboratory of Ultra-intense Laser Science and Technology, Shanghai Institute of Optics and Fine Mechanics, Chinese Academy of Sciences, Shanghai 201800, People’s Republic of China}
	\affiliation{Institut f\"{u}r Theoretische Physik I, Heinrich-Heine-Universit\"{a}t D\"{u}sseldorf, 40225 D\"{u}sseldorf, Germany}

    \author{Dimitris Sofikitis}
    \affiliation{Department of Physics, Atomic and Molecular Physics Laboratory, University of Ioannina, University Campus, 45110 Ioannina, Greece}

    \author{Oliver Mathiak}
    \affiliation{Institut f\"{u}r Theoretische Physik I, Heinrich-Heine-Universit\"{a}t D\"{u}sseldorf, 40225 D\"{u}sseldorf, Germany}

    \author{T. Peter Rakitzis}
    \affiliation{Institute of Electronic Structure and Lasers, Foundation for Research and Technology-Hellas, 71110 Heraklion-Crete, Greece}
    \affiliation{Department of Physics, University of Crete, 70013 Heraklion-Crete, Greece}

    \author{Bernhard Hidding}
	\affiliation{Institut f\"{u}r Laser- und Plasmaphysik, Heinrich-Heine-Universit\"{a}t D\"{u}sseldorf, 40225 D\"{u}sseldorf, Germany}

	\author{Alexander Pukhov}
	\affiliation{Institut f\"{u}r Theoretische Physik I, Heinrich-Heine-Universit\"{a}t D\"{u}sseldorf, 40225 D\"{u}sseldorf, Germany}

    \author{Liangliang Ji}
    \affiliation{State Key Laboratory of Ultra-intense Laser Science and Technology, Shanghai Institute of Optics and Fine Mechanics, Chinese Academy of Sciences, Shanghai 201800, People’s Republic of China}

    \author{Markus B\"{u}scher}
	\affiliation{Peter Gr\"{u}nberg Institut (PGI-6), Forschungszentrum J\"{u}lich, 52425 J\"{u}lich, Germany}
	\affiliation{Institut f\"{u}r Laser- und Plasmaphysik, Heinrich-Heine-Universit\"{a}t D\"{u}sseldorf, 40225 D\"{u}sseldorf, Germany}
    
	\date{\today}
	
	\begin{abstract}
	Pinching of the driver beam in plasma wakefield acceleration is generally considered an unwanted effect that needs to be mitigated. Here, we propose that this effect can be utilized for the injection of spin-polarized electron beams from hydrogen halide targets into wakefields. Particle-in-cell simulations show that the electron spin is preserved on a level of 50\% for a wide range of parameters due to the injection geometry. The presented injection scheme provides a possible pathway to alleviate some of the restrictions associated with pre-polarized hydrogen halide targets. 
	\end{abstract}
	
	\maketitle

    \section{Introduction}
    Plasma-based acceleration schemes have been shown to be feasible alternatives to conventional rf-based approaches, as they enable stronger accelerating gradients. Currently, high-energy electron beams on the 10 GeV-level \cite{Picksley2024}, with ultra-low emittances \cite{Hidding2012} can be produced. 
    Based on the P5 Report in 2023 \cite{snowmass}, one overarching goal for the community in the long term will be to pave the way for a 10 TeV pCM collider \cite{Gessner2025}.
    In such colliders, the spin polarization of the interacting beams can become important \cite{MoortgatPick2008}. Moreover, polarization of probe beams is highly relevant for deep-inelastic scattering experiments to understand the nucleon structure \cite{Glashausser1979}.
    Recently, numerous papers studying the production and acceleration of polarized lepton and ion beams from laser-plasma interaction have been published. In contrast to conventional polarized sources like Ref. \cite{Litvinenko2026}, plasma-based sources could potentially provide higher currents and allow the realization of high-energy polarized beams in laboratories without access to GaAs photocathodes. A general overview of the field is given by Reichwein \textit{et al.} \cite{Reichwein2025}.
    While a first proof-of-principle experiment on the acceleration of nuclear-polarized helium-3 has been performed \cite{Zheng2026}, no experimental results on polarized electrons from wakefields have been produced thus far. However, projects like ``Laser Electron Acceleration with Polarization'' (LEAP) at DESY are on-going \cite{Stehr2025}.

    Particle-in-cell (PIC) simulation studies by Wu \textit{et al.} first showed the feasibility of both PWFA and LWFA for polarized electron beams, in case a pre-polarized target was utilized \cite{Wu2019lwfa, Wu2019pwfa}. Here, it was found that the injection into the wakefield is particularly crucial, as the interaction of the electron spin with the strong fields inside the cavity induces significant spin precession. In contrast, the acceleration phase where $\gamma \gg 1$, only changes the spin marginally, as the precession frequency is reduced by a factor $\sim \gamma^{-1}$.
    For a fully pre-polarized target Wu \textit{et al.} demonstrated a final witness polarization of $\sim 80\%$.
    In two studies, Nie \textit{et al.} proposed in-situ setups that rely on the ionization of specific orbitals like $4f^{14}$ for Yb and do not require pre-polarization \cite{Nie2021, Nie2022}. Such schemes yield up to 4 kA beams with approx. 56\% polarization.

    One difficulty in realizing many of the proposed schemes are restrictions in the available target parameters. As discussed by Sofikitis \textit{et al.} \cite{Sofikitis2025}, hydrogen halide targets (which make up most of the proposed setups) only provide a comparatively low density and a small interaction volume.
    These considerations are often neglected, instead simulating a broader parameter regime for which, currently, no significant degree of pre-polarization could be produced. For example, based on the calculations in \cite{Sofikitis2025}, for achieving 90\% pre-polarization at $10^{17}$ cm$^{-3}$, only a target diameter of up to 1 {\textmu}m can be achieved.

    In this paper, we propose to utilize the pinching of the driving electron beam during propagation for specific ionization of the polarized electrons. Effects like scalloping have long been known in the context of wakefield accelerators \cite{Hidding2012, Blumenfeld2007}, but are generally considered as unwanted as they can be detrimental to the witness quality.
    Beam-induced ionization injection has been considered as a potential mechanism in the past \cite{Oz2007, Vafaei2019}, however, without considering spin polarization and the specific target requirements of hydrogen halides.
    Here we show that pinching injection can be utilized in coalition with a pre-polarized hydrogen halide target to provide high-energy, polarized ($P \sim 50\%$) electron sources.

    \begin{figure*}[ht]
        \centering
        \includegraphics[width=\textwidth]{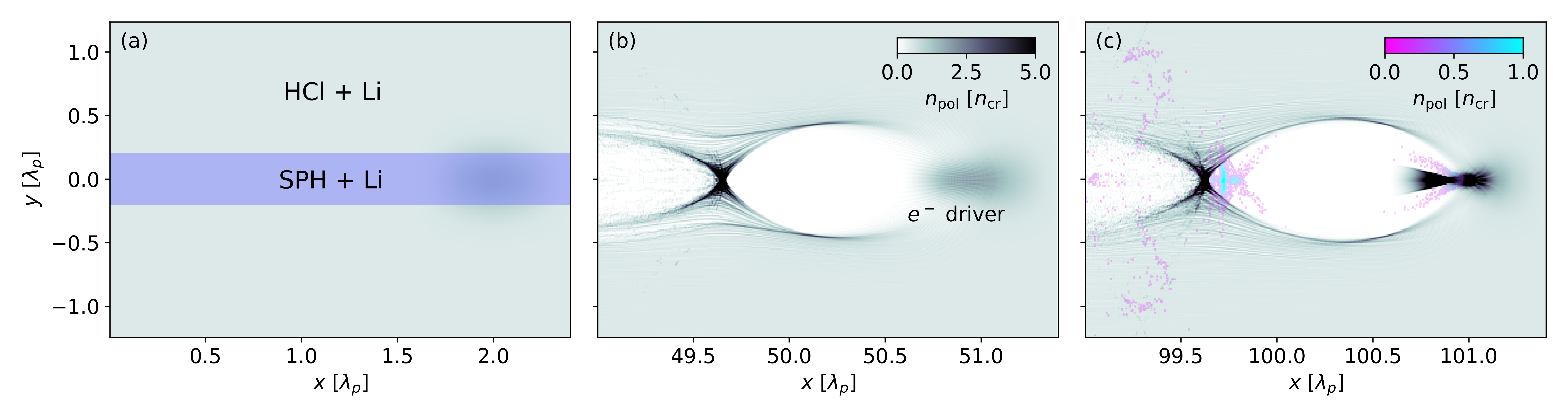}
        \caption{\label{fig:scheme} Depiction of the pinching injection scheme. (a) The target consists of a broad Li + HCl component (gray) and a narrow channel of spin-polarized hydrogen (SPH, blue). (b) The electron driver generates a wakefield in the lithium component while leaving the SPH non-ionized. (c) Upon pinching of the mismatched driver, the high-density region can ionize the SPH component, leading to the injection of spin-polarized electrons (magenta/blue) into the wake.}
    \end{figure*}

    \section{Simulation setup}

    We conduct 3D-PIC simulations using the code \textsc{vlpl} \cite{Pukhov1999, Pukhov2016} with a simulation domain of size $2.5 \lambda_p \times 2.5 \lambda_p \times 2.5 \lambda_p$ with $\lambda_p = 300$ {\textmu}m. The grid resolution is chosen as $h_x = 0.005 \lambda_p$ ($x$ being the propagation direction), $h_y=h_z=0.02\lambda_p$ with a time step of $\Delta t = h_x / c$, in accordance with the rhombi-in-plane solver \cite{Pukhov2020}.

    The scheme works as follows (cf. Fig. \ref{fig:scheme} for a visualization): (a) The plasma target consists of an HCl component and an additional Li component. The lithium is used as a low ionization-threshold medium, while the spin-polarized hydrogen (SPH) acts as a high ionization-threshold medium.
    The lithium component is modeled as a homogeneous slab of density $0.5 n_0$ throughout the whole simulation domain, with $n_0 = m_e \omega_p^2 / (4\pi e^2) \approx 1.2 \times 10^{16} \; \mathrm{cm}^{-3}$.  The slab has an initial up-ramp and is pre-ionized to the state Li$^{1+}$ throughout. 
    The HCl is initially non-ionized and modeled as a super-Gaussian channel ($n(r) \propto 1 - \exp(-r^8/r_c^8)$) with a radius $r_c = 0.2\lambda_p$, and peak density of $0.5 n_0$. A comparison of simulations with a distinct Cl component and without (i.e. just H $+$ Li) shows that the presence of Cl affects simulation results only marginally, so the chlorine component is ignored subsequently to reduce computational effort. As extensively discussed by Sofikitis \textit{et al.} in Ref. \cite{Sofikitis2025}, a wide range of densities for the hydrogen halide target can theoretically be chosen. This, however, determines the maximal size of the interaction volume as well as the maximal degree of pre-polarization that can be achieved during preparation.

    (b) A Gaussian-shaped electron beam with $\sigma_{x,y,z} = 0.17 \lambda_p$ and a density of $1 n_0$ drives a wakefield in the lithium component of the target while leaving the SPH unaffected. The driver has a momentum spread of $\sigma_{p_x} \approx 0.15 m_e c$, $\sigma_{p_y} = \sigma_{p_z} \approx 0.011 m_e c$ and a mean energy of (1 - 10) GeV, depending on the simulation run.
    At $t=0$, the driver is centered around $x = 2 \lambda_p$ in the co-moving simulation window. 

    (c) As the driver propagates through the plasma, it will start to pinch which effectively increases the on-axis driver density. In turn, the electromagnetic fields induced by the pinched beam can ionize the SPH. Similarly to the original approaches using hydrogen halide targets \cite{Sofikitis2025, Wu2019lwfa, Wu2019pwfa}, ionization in this parameter regime should only marginally affect the initial electron polarization (cf. Ref. \cite{Klaiber2014} for a discussion focusing on ionization due to strong laser fields). One crucial requirement for our setup (as is the case for \cite{Sofikitis2025}) is that the ionization is timed in accordance with the hyperfine oscillations by which the polarization is transferred periodically between the hydrogen nuclei and the electrons. In the case of our simulations, we assume that the electrons from the SPH channel are initially fully polarized $(s_x \equiv +1)$. The electrons will fall back in the wakefield while gaining longitudinal momentum until they are trapped.
    After ionization, electron spin precesses according to the T-BMT equation \cite{Thomas1926, Bargmann1959},
         \begin{align}
        \frac{\D \ssb}{\D t} = - \Omegab \times \ssb \; ,
    \end{align}
    where
    \begin{align}
        \Omegab = \frac{e}{mc} \left[ \Omega_B \Bb - \Omega_v \left( \frac{\vb}{c} \cdot \Bb \right) \frac{\vb}{c} - \Omega_E \frac{\vb}{c} \times \Eb \right] \; , \label{eq:prec}
    \end{align}
    is the precession frequency with $e$ denoting the charge, $m$ the mass, $c$ the vacuum speed of light, $\vb$ the particle velocity and $\Eb, \Bb$ the electromagnetic field. The prefactors are defined as
    \begin{align}
        \Omega_B = a + \frac{1}{\gamma} \; , && \Omega_v = \frac{a \gamma}{\gamma + 1} \; , && \Omega_E = a + \frac{1}{\gamma + 1} \; ,
    \end{align}
    with the anomalous magnetic moment $a$. For electrons, we have $a_e \approx 10^{-3}$.
    In our parameter regime, other effects like the Stern-Gerlach force or radiative processes like the Sokolov-Ternov effect can be neglected \cite{Thomas2020}.

    \section{Results \& Discussion}
\begin{figure}[h]
        \centering
        \includegraphics[width=0.5\textwidth]{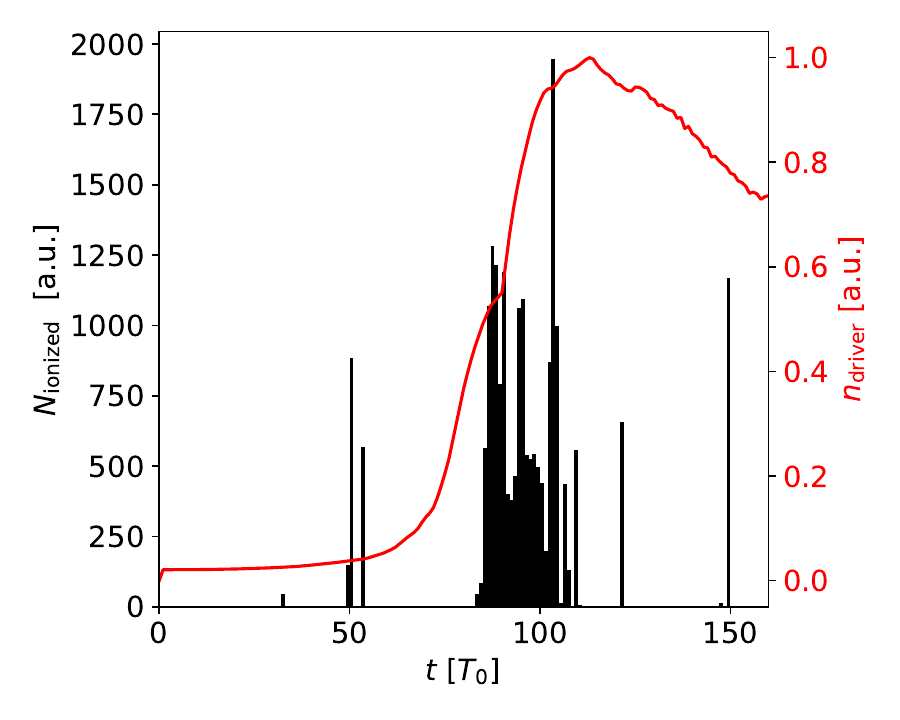}
        \caption{Number of \textit{tracked} electrons ionized from the hydrogen component per time interval (black histogram). Note that only every 1000th electron in that particle species was tracked. The red line corresponds to the maximum driver density at any given time.}
        \label{fig:inj_time}
    \end{figure}

    The PIC simulations show that initially, when the driver is not yet deformed, ionization of SPH only occurs towards the high-density region at the bubble tail.
    After sufficient propagation through the plasma, the driver will start to pinch, leading to fields at the front of the wake that are capable of field ionizing some of the SPH (see Fig. \ref{fig:scheme}(c)). 
    The onset of ionization events can be seen in Fig. \ref{fig:inj_time}. Here, the histogram (black) shows the number of ionized electrons at a given time. Clearly, the ionization is due to the pinching of the driver (cf. red line for the maximum driver density).
    Since these (polarized) electrons need to traverse the whole cavity, instead of only the bubble tail, they gain significant longitudinal momentum, allowing them to be trapped.

\begin{figure}[h]
        \centering
        \includegraphics[width=0.5\textwidth]{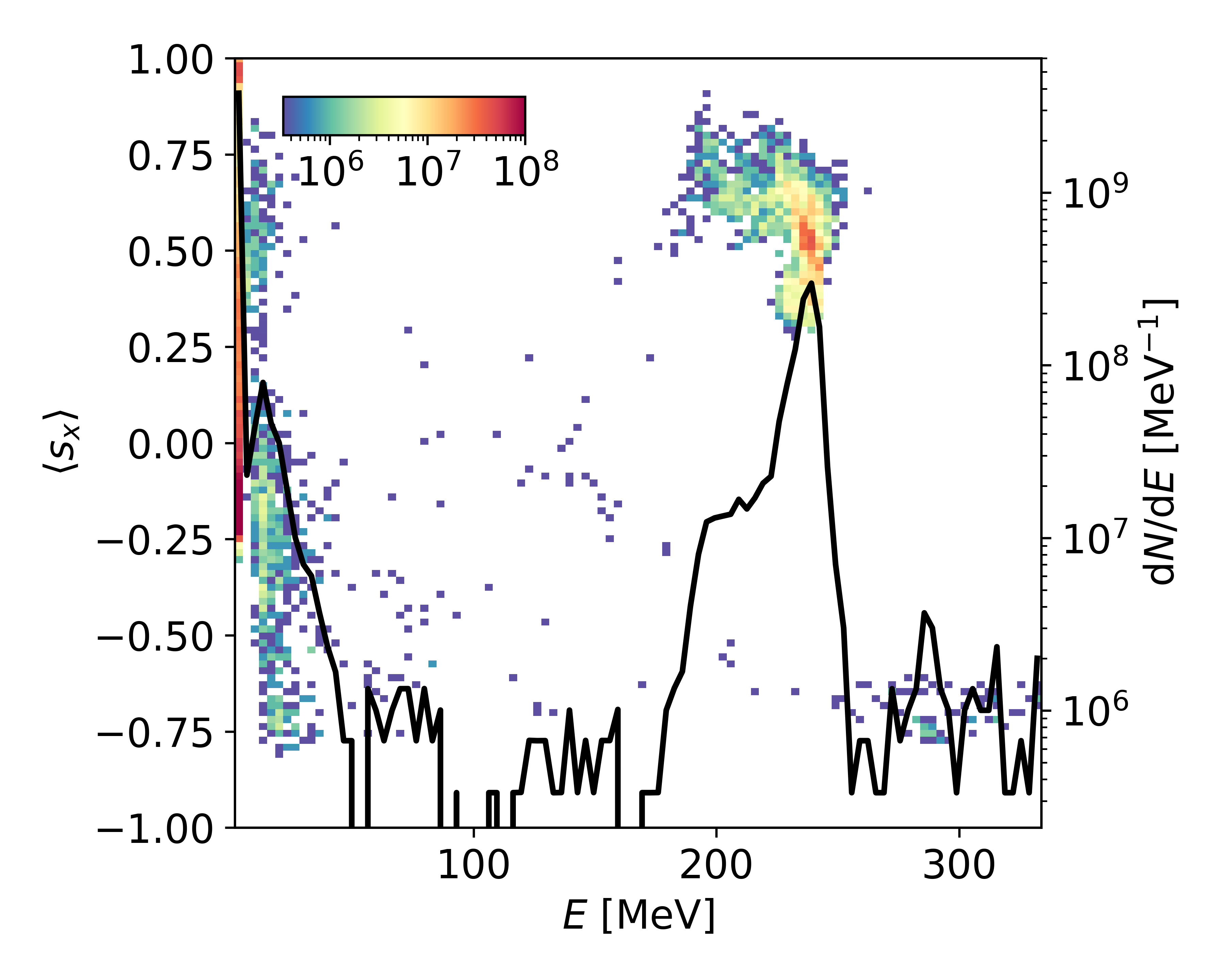}
        \caption{\label{fig:spectrum} Electron energy spectrum (black line) after $160 T_0$. The heatmap shows the number of particles in $E$-$s_x$ space.}
    \end{figure}

    When the electrons are falling back in the wake, the spin is subject to the fields inside the bubble. As discussed in Ref. \cite{Kostyukov2004}, they can be approximated as
        \begin{align}
        E_x = \frac{\xi}{2} \; , && E_y = - B_z = \frac{y}{4} \; , \notag \\
        B_x = 0 \; , && E_z = B_y = \frac{z}{4} \; .
    \end{align}
    Previous studies, e.g. \cite{Wu2019lwfa,Wu2019pwfa}, have shown that at low energies, strong transverse fields can be a limiting factor for beam polarization.
    Since in our scheme the pinching occurs close to the optical axis, meaning that the injected electrons are predominantly born close to $y = z = 0$, these detrimental transverse field contributions essentially vanish.
    For the 10 GeV driver, we observe a well-defined peak in the energy spectrum (cf. Fig. \ref{fig:spectrum}) and witness polarization of approx. 51\%.
    Varying the driver energy changes the time at which the pinching occurs. Higher driver energy leads to delayed pinching, meaning that the electrons are injected and accelerated later. Accordingly, we observe the lowest witness energy (peak at 235 MeV after 160 $T_0$) for a 10 GeV driver, whereas the highest energy (437 MeV) is obtained for 1 GeV. Beam charge increases from 65 pC for the 1 GeV driver to approx. 187 pC for the 10 GeV driver. 
    
\begin{figure}[h]
        \centering
        \includegraphics[width=0.5\textwidth]{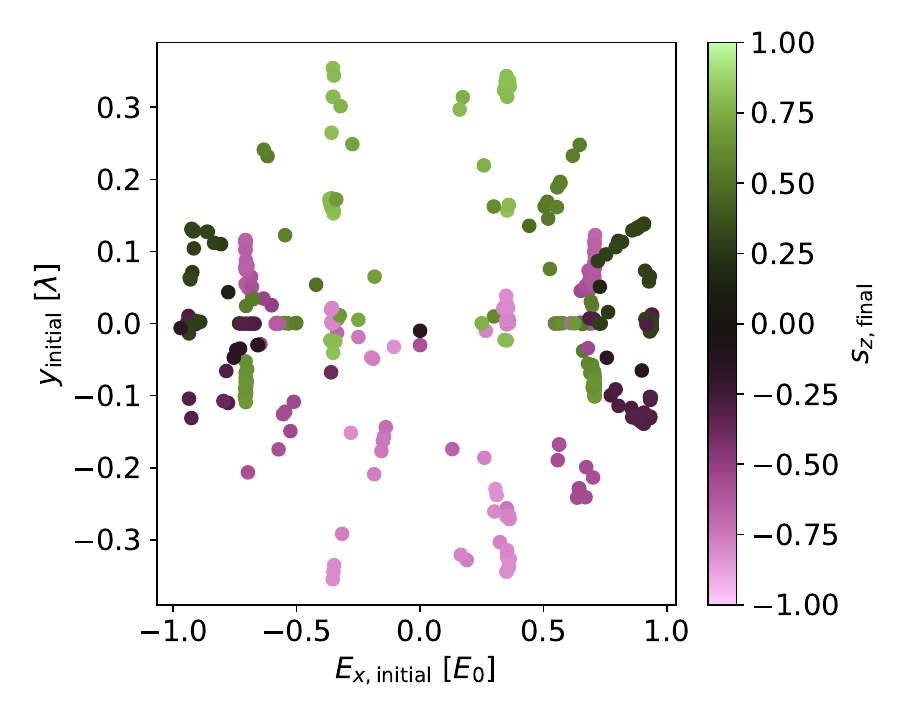}
        \caption{Initial longitudinal field $E_x$ plotted against the initial $y$-position of the high-energy witness beam electrons. The color coding denotes the final value of the spin component $s_z$. A clear distinction between upper and lower half-plane can be made, with the distinct lines at $E_x \approx \pm 0.75 E_0$ stemming from a strong ionization event with local field inversion.}
        \label{fig:spin_scatter}
    \end{figure}

    In any case, the polarization remains around 50\%. 
    This value remains rather stable, since the polarization yield is based on the radial symmetry of the injection process. Looking at the different terms of the precession in frequency in Eq. \eqref{eq:prec}, we observe that the spin rotation in $z$-direction, $\D s_z / \D t$, is dependent on the transverse position. Thus, we expect particles in the upper half-plane to have a counter-rotating spin motion compared to particles from the lower half-plane. This assumption is verified when plotting the initial $E_x$ experienced by the injected particles against their initial $y$-position and the final, transverse spin component $s_z$ (cf. Fig. \ref{fig:spin_scatter}). 

    Consistently, particles initially located at $y > 0$ end up with $s_z > 0$, while particles with $y < 0$ have spin $s_z < 0$. There are, however, two well-defined lines with seemingly inverted spin behavior around $E_x \approx \pm 0.75 E_0$. These particle clouds can be attributed to distinct pinching events (thus having the same initial $E_x$), where the pinching leads to a localized inversion of $E_x$ compared to the rest of that bubble half.

    The consistent polarization of approx. 50\% is also observed when modifying our geometry: in separate simulations, we displace the 10 GeV driver transversely by $(0.1 - 0.4)\lambda_p$, but still obtain comparable maximum energy and beam charge. Polarization increases slightly to 57\% for the strongest offset, indicating further that the value is based on the injection geometry. Similarly, polarization remains around 50\% for a narrower ($0.1\lambda_p$) SPH channel. Therefore, the pinching injection scheme could enable the realization of a wakefield-based, polarized electron source despite the restrictions of the hydrogen halide targets. The mechanism is technically not limited to the driver energies presented here, but could also be realized using lower-energy drivers. Future improvements to the proposed scheme could consist in employing a spin filter like \cite{Wu2020}, where the polarization could be increased from 35\% to 80\%.
    \section{Conclusion}
    We have presented a injection scheme for spin-polarized electrons from hydrogen halide targets that utilizes the pinching of a mismatched driver beam.
    The high-density region of the pinched driver is capable of ionizing electrons from the pre-polarized hydrogen component. Since the ionization occurs close to the optical axis, the electrons are subject to only weaker transverse fields, limiting depolarizing effects during injection.
    The polarization of the witness beam remains stable at approx. 50\% for a wide variety of parameters due to the injection geometry. Since the pre-factors of the precession frequency scale as $\Omega \propto \gamma^{-1}$, the electrons could be further accelerated in subsequent stages without affecting the polarization.
    Moreover, since the ionization is highly localized based on the driver parameters, this method relieves the restrictions currently imposed on other means of accelerating electrons from pre-polarized sources \cite{Sofikitis2025}.

    \begin{acknowledgments}
    The authors gratefully acknowledge the Gauss Centre for Supercomputing e.V. \cite{GCS} for funding this project (spaf) by providing computing time through the John von Neumann Institute for Computing (NIC) on the GCS Supercomputer JUWELS at J\"ulich Supercomputing Centre (JSC). The work of M.B. has been carried out in the framework of the JuSPARC (J\"ulich Short-Pulse Particle and Radiation Center \cite{JuSPARC}). 
    \end{acknowledgments}

    \bibliography{pinch_bib}

\end{document}